# Modelling the energy gap in transition metal/aluminium bilayers

G. Brammertz[1], A. Golubov[2], A. Peacock[1], P. Verhoeve[1], D. J. Goldie[3], R.Venn[4].

[1] *Astrophysics Division, Space Science Department, European Space Agency, ESTEC, Noordwijk, The Netherlands*

[2] *Faculty of Applied Physics, Low Temperature Group, University of Twente, Enschede, The Netherlands*

[3] *Oxford Instruments Scientific Research Division, Newton House, Cambridge Business Park, Cowley Road, Cambridge CB4 4WZ, United Kingdom*

[4] *Cambridge MicroFab Ltd., Trollheim Cranes Lane, Kingston, Cambridge CB3 7NJ, United Kingdom*

**Abstract**

We present an application of the generalised proximity effect theory. The theory has been used to determine the energy gap ($\Delta_g$) in proximised transition metal – aluminium bilayer structures such as Nb/Al, Ta/Al, V/Al and Mo/Al. These bilayers have different film thicknesses ranging from 5 to 260 nm. For the cases of Nb/Al, Ta/Al and V/Al bilayers, the interface parameters $\gamma$ and $\gamma_{BN}$ (here we define $\gamma$ as the ratio of the products of normal state resistivity and coherence length in each film of the bilayer while $\gamma_{BN}$ is the ratio of the boundary resistance between film 1 and 2 to the product of the resistivity and coherence length in the second film), which were used as input parameters to the model, were inferred experimentally from an existing bilayer of each kind and then suitably modified for different film thicknesses. This experimental assessment is therefore based on a comparison of measurements of the critical temperature and the energy gap at 300mK with the predictions from the model for various values of $\gamma$, $\gamma_{BN}$. The energy gap of the bilayer was experimentally determined by using symmetrical Superconducting Tunnel Junctions (STJs) of the form S-Al-AlOx-Al-S, where each electrode corresponds to a proximised bilayer. However for the case of Mo/Al bilayers the interface parameters were determined theoretically since currently no STJ data for this configuration are available. The results for the Nb/Al, Ta/Al and V/Al bilayers have also then been compared to experimentally determined energy gaps found for a series of STJs with different film thicknesses. The correspondence between experiment and theory is very good.

# I. Introduction

The deposition of a superconductor $S_1$ on a superconductor $S_2$ modifies the properties of both $S_1$ and $S_2$ materials. Predictions of the value of the energy gap in such superconducting bilayer films are important for applications such as Superconducting Tunnel Junctions (STJs) [1] and Transition Edge Sensors (TES) [2]. Recent developments in the proximity effect theory [3] now allow a systematic application of the model to a wide variety of bilayers having arbitrary film thicknesses.

In this work the model was applied to transition metal/aluminium bilayers such as Ta/Al, Nb/Al, V/Al and Mo/Al.

Section II presents a summary of the salient features of the proximity effect theory and the necessary input parameters to the model. Section III describes in detail the specific input parameters and their dependence on the film thickness. The determination of the key parameters describing the $S_1$-$S_2$ interface is also described and both theoretical and experimental determinations are discussed. In section IV the results of the numerical simulations are presented and compared to experimental values found for STJs with different geometrical/material lay-ups.

# II. The proximity effect model

The proximity effect theory has recently been generalised to the case of arbitrary $S_1$ and $S_2$ layer thicknesses in the most general case of a finite critical temperature for the $S_2$ layer [3]. For infinite films in the y and z direction and with the film surface perpendicular to the x-axis it was shown that the Usadel equations [4] describing the quasiparticle density of states could be written as

$$\xi_{S_i}^2 \theta_{S_i}''(x) + i\varepsilon \sin\theta_{S_i}(x) + \Delta_{S_i}(x)\cos\theta_{S_i}(x) = 0, \qquad (1)$$

where the pair potential $\Delta_{S_i}$ is determined by the self-consistency relation

$$\Delta_{S_i}(x)\ln\frac{T}{T_{c,S_i}} + 2T\sum_{\omega_n}\left[\frac{\Delta_{S_i}(x)}{\omega_n} - \sin\theta_{S_i}(i\omega_n, x)\right]. \qquad (2)$$

The function $\theta_{S_i}$ is a unique Green's function, which defines the quasiparticle density of states $N_{S_i}$ according to the relation:

$$\frac{N_{S_i}(\varepsilon, x)}{N_{S_i}(0)} = \operatorname{Re}[\cos\theta_{S_i}(\varepsilon, x)]. \qquad (3)$$

Here $\omega_n$ is the Matsubara frequency, which is related to the quasiparticle energy $\varepsilon$ by the relation $\omega_n = -i\varepsilon$, $\xi_{S_i}$ is the coherence length and $N_{S_i}(0)$ is the electronic density of states in the normal state at the Fermi surface.

In a bilayer comprising two different superconducting layers $S_1$ and $S_2$ the Usadel equations (1) and (2) must be solved in every layer with the use of the appropriate boundary conditions. Here the origin of the coordinate system is chosen at the $S_1$-$S_2$ interface and the region with x>0 refers to the $S_1$ layer while x<0 is for the $S_2$ film, where $S_2$ is the low energy gap material. The film thicknesses are respectively $d_{S_1}$ and $d_{S_2}$. At the $S_1$-$S_2$ interface the boundary conditions are [5]:

$$\gamma_{BN}\xi_{S_2}^*\theta_{S_2}' = \sin(\theta_{S_1} - \theta_{S_2}) \qquad (4)$$

$$\gamma\xi_{S_2}^*\theta_{S_2}' = \xi_{S_1}\theta_{S_1}', \qquad (5)$$

where $\xi_{S_2}^* = \xi_{S_2}\sqrt{\frac{T_{c,S_2}}{T_{c,S_1}}}$ is the coherence length in layer $S_2$ normalised to the critical temperature of layer $S_1$. $\gamma$ and $\gamma_{BN}$ are the interface parameters describing the nature of the interface between the two materials. They are defined by:

$$\gamma = \frac{\rho_{S_1}\xi_{S_1}}{\rho_{S_2}\xi_{S_2}^*} \quad (6)$$

$$\gamma_{BN} = \frac{R_B}{\rho_{S_2}\xi_{S_2}^*}. \quad (7)$$

Here $\rho_{S_1}$ and $\rho_{S_2}$ are the normal state resistivities and $R_B$ is the product of the resistance of the $S_1$-$S_2$ boundary and its interface.

At the free interface of both $S_1$ and $S_2$ layers the boundary conditions are

$$\theta'_{S_1}(d_{S_1}) = 0 \quad (8)$$

$$\theta'_{S_2}(-d_{S_2}) = 0. \quad (9)$$

## III. The model's input parameters

The bilayer characteristics that are required to apply the proximity effect model are the individual monolayer critical temperatures and coherence lengths and the interface parameters $\gamma$ and $\gamma_{BN}$. In order to be able to apply the model to bilayers with different film thicknesses we initially establish the dependence of these parameters on the monolayer film thickness.

### A. Input parameters as a function of film thickness

#### 1. Critical temperature

The critical temperature of a thin film can be determined according to the model of Cooper [6]. This model states that superconductivity is lost in a thin surface layer of thickness $b_t$ due to a reduction in the electron density of states near the surface. The calculated critical temperature dependence according to such a model is:

$$T_c = T_{c,0}\left(1 - \frac{b_t}{N\nu t}\right),$$
(10)

where $N$ is the electron density of states at the Fermi level, $\nu$ the bulk interaction potential and $t$ the film thickness. The factor 2 has been omitted in this formula since the thin layer is only bound on one side by a non-superconducting layer.

#### 2. Coherence length

In the dirty limit ($\Lambda < \xi$) the coherence length is given by:

$$\xi = \sqrt{\frac{\xi_0 \Lambda}{3}},$$
(11)

where $\xi$ is the coherence length in the film, $\xi_0$ is the bulk coherence length and $\Lambda$ is the mean free path in the film.

In the clean limit ($\Lambda > \xi$) the coherence length is found using:

$$\xi = \left(\frac{1}{\xi_0} + \frac{1}{\Lambda}\right)^{-1}.$$
(12)

From this we can write the normalised coherence length in the second film:
$$\xi_2^* = \xi_2 \sqrt{\frac{T_{c,2}}{T_{c,1}}}.$$
(13)

## 3. Mean free path

The bulk mean free path at 10K, $\lambda_{10}$, can be found via the bulk mean free path at 300K, $\lambda_{300}$, and the residual resistance ratio ($RRR_b$) of a 'thick' film (equivalent to the bulk material):

$$\lambda_{10} = RRR_b\, \lambda_{300}.$$
(14)

Once the bulk mean free path at 10K known, we can calculate the mean free path $\Lambda$ in a thin film ($t \ll \lambda_{10}$), using [7]:

$$\Lambda = \frac{3t}{4}\left(\ln\frac{\lambda_{10}}{t} + 0.423\right),$$
(15)

where t is the film thickness.

When this approximation does not apply, we can still determine the mean free path by using the following equation, which is valid for larger film thicknesses [8]:

$$\Lambda = \lambda_{10} - \frac{3\lambda_{10}^2}{8t}(1-p),$$
(16)

where p is the proportion of elastically reflected electrons at the film interfaces. A value of p = 0.5 was found to be a reasonable approximation in thin film experiments. In practice equation (15) is applied when $t/\lambda_{10} < 0.3$.

## 4. Interface parameters

The biggest problem in developing the model and comparing predictions with experimental data is in the determination of the interface parameters described by (6) and (7). Substituting equations (11) and (13) into equation (6), we find:

$$\gamma = C_\gamma \sqrt{\frac{\Lambda_2 T_{c,1}}{\Lambda_1 T_{c,2}}},$$
(17)

with
$$C_\gamma = \frac{\rho_1 \Lambda_1}{\rho_2 \Lambda_2} \sqrt{\frac{\xi_{0,1}}{\xi_{0,2}}}.$$
(18)

The Drude free electron model states that the product $\rho\Lambda$ is a material constant, which can be written as:

$$\rho \Lambda = \frac{m v_F}{e^2 n},$$
(19)

where $v_F$ is the Fermi velocity, n the number of free electrons per unit volume and m is the effective mass of a conduction electron. From this it follows that $C_\gamma$ is a constant that depends only on the nature of the two materials involved.

Following the same approach with equation (7) we obtain:

$$\gamma_{BN} = C_{\gamma_{BN}} \sqrt{\frac{T_{c,1} \Lambda_2}{T_{c,2}}},$$
(20)

with 
$$C_{\gamma_{BN}} = \frac{R_B}{\rho_2 \Lambda_2} \sqrt{\frac{3}{\xi_{0,2}}}.$$
(21)

Here $C_{\gamma_{BN}}$ is a factor that depends strongly on the microscopic structure of the interface between the two films via the factor $R_B$. Note we assume that this interface is the same for any bilayer comprising the same two materials.

## B. Basic parameters

In order to calculate input parameters to the model from equations (10) to (21) we need to establish a number of basic parameters for all of the materials. These are summarised in table 1 for all 5 materials discussed in this paper.

The values in the first three columns were established from experiments on thin films in our labs. The values for Nv can be found in reference [9]. The values for the mean free path at 300K were taken from different references [3,10,11].

## C. Determination of the interface constants

Here we determine the interface constants $C_\gamma$ and $C_{\gamma_{BN}}$, which are constant with respect to film thickness. The simplest approach to this determination is by using theoretical values for all quantities appearing in equations (18) and (21). This is the only realistic approach when no experimental data on a particular type of bilayer exists. When however experimental values of the critical temperature and the energy gap of just one bilayer are available, we can determine the interface parameters for this bilayer experimentally. This is performed by comparing the measured energy gap and critical temperature to those values calculated with the model described in section II for different interface parameters. We can then deduce the interface constants from the values of the interface parameters for the particular bilayer of interest by using equations (17) and (20). The resultant values of these interface constants can then be used for all possible lay-ups of a specific bilayer configuration.

This experimental approach has been adopted for Ta/Al, Nb/Al and V/Al bilayers. The critical temperature and energy gap were deduced from experiments on Superconducting Tunnel Junctions (STJs) of the form S-Al-AlOx-Al-S, where S is either Ta, Nb or V. For all three junctions the energy gap values were determined at 300 mK. Figures 1(a-c) show that the two values (critical temperature and energy gap at 300 mK) are adequate to ensure an exact determination of both interface parameters. In figure 1 the two lines represent possible combinations of the interface parameters that satisfy the experimentally determined critical temperature or energy gap. For all these three bilayers there is a single unique solution for the pair of interface parameters.

The Ta/Al STJ used for the determination of the interface parameters is based on a bilayer with a Ta film thickness of 100nm and an Al film thickness of 55nm. The measured energy gap at 300 mK and critical temperature of this STJ are 0.45 meV and 4.42 K respectively. From this data we derive $\gamma = 0.13$ and $\gamma_{BN} = 5.5$ (Fig. 1a), implying interface constants $C_\gamma$ and $C_{\gamma_{BN}}$ of 0.0884 and 0.395 nm$^{-1/2}$ respectively.

The Nb/Al STJ has a Nb film thickness of 100 nm and an Al film thickness of 30 nm. The energy gap at 300 mK and critical temperature are 0.85 meV and 8.55 K, respectively leading to $\gamma = 0.595$ and $\gamma_{BN} = 3$ (Fig 1b) with interface constants $C_\gamma = 0.397$ and $C_{\gamma_{BN}} = 0.179$ nm$^{-1/2}$. These values are in reasonable accordance with previous experimental determinations of the interface parameters for Nb/Al STJs with different Al film thicknesses by Zehnder et al. [12]. In this reference the derived values for a Nb/Al STJ with 200 nm of Nb and 30 nm of Al were $\gamma = 0.82$ and $\gamma_{BN} = 3$. Note that the difference in the $\gamma$ value can be explained by the different Nb film characteristics. Whereas Zehnder et al. report RRR values of 1.7 for polycrystalline Nb, our base film RRR values are of the order of 90, typical values for epitaxial Nb films. Dmitriev et al. [13] also determined the values of the interface parameters for Nb/Al bilayers experimentally, but without taking the film thickness dependence into account. They derived the values $\gamma = 0.3$ and $\gamma_{BN} = 1$.

The V/Al STJ has symmetrical electrodes with 100 nm of V and 8 nm of Al. The critical temperature is equal to 5.12 K and the energy gap is 0.65 meV at 300 mK. The interface parameters $\gamma$ and $\gamma_{BN}$ are 0.33 and 2.95 respectively (Fig 1c), with interface constants $C_\gamma$ and $C_{\gamma_{BN}}$ of 0.29 and 0.39 nm$^{-1/2}$ respectively.

For Mo/Al bilayers, where no Mo-based STJ's have yet been fabricated a theoretical approach to the determination of the interface parameters was adapted. Estimates from the literature of all quantities appearing in equations (18) and (21) were used to yield values of $C_\gamma = 0.33$ and $C_{\gamma_{BN}} = 0.39$ nm$^{-1/2}$. For the interface resistance $R_b$ it should be stressed that the same value as for V/Al interfaces was assumed since this has the closest lattice match to the Mo/Al system.

## IV. Results of the simulations

### A. General results

In this section we will present the general results obtained from the model. For every type of bilayer 16 different lay-ups are presented where the thickness of each film takes the values 5, 20, 50 and 100 nm. For each lay-up the quasiparticle density of states was calculated throughout the bilayer from which the energy gap could be derived. For the Ta/Al, Nb/Al and V/Al bilayers all calculations were performed at a temperature of 300 mK. For the Mo/Al bilayer having a significantly lower $T_c$ the temperature was taken as 20 mK. All input parameters were calculated from the formulas presented in section III A, with the basic parameter values presented in sections III B and C.

## 1. Ta/Al, Nb/Al and V/Al bilayers

The behaviour for Ta/Al, Nb/Al and V/Al bilayers is very similar and we therefore treat these cases together. Table 2 provides the calculated energy gap at 300 mK as a function of film thicknesses for Ta/Al, Nb/Al and V/Al bilayers while figures 2, 3 and 4 illustrates the overall shape of the energy gap surface as a function of bilayer geometry. As expected the energy gap increases with increasing transition metal thickness and of course decreases with increasing Al film thickness. Figures 5, 6 and 7 illustrate the calculated density of states through a bilayers having a 100 nm film of the transition metal (Ta, Nb or V) together with 50 nm of Al. Here (a) represents the density of states at four key positions in the bilayer. The dashed lines represent the density of states at the $S_1$-Al interface, in the $S_1$ and in the Al, while the solid lines provide the density of states at the free interfaces in both films. Figures 5(b) to 7(b) illustrate the evolution of the density of states surface as a function of both the position through the whole bilayer and energy. In the transition metal films the density of states peaks at the bulk energy gap of the transition metal. Below this value only few states are available compared to the number of states available in the Al at these energies. This actually implies a high number of Andreev reflections at the interface. In the Al most states are however available up to the overall bilayer energy gap, with a second peak at the transition metal bulk energy gap.

## 2. Mo/Al bilayers

The behaviour for Mo/Al bilayers is somewhat different. Table 3 gives the energy gap as a function of film thickness while figure 8 illustrates the energy gap surface as a function of bilayer geometry. Clearly from figure 8 Mo is the predominant superconductor in the bilayer. This is due to its very large mean free path at 10K, which induces a very large coherence length. Note that for thicker Mo films (100 nm) the energy gap is independent of the Al thickness and equal to the value of bulk Mo (0.140 meV). For thinner Mo films (5 nm) the influence of the Al becomes apparent although, but even for an Al thickness of 100nm with only 5 nm of Mo, the energy gap is increased to only 0.148 meV. Only for very large thicknesses of the Al film the overall bilayer gap converges to the value of bulk Al (1.5 μm of Al: 0.165eV; 33 μm of Al: 0.180 meV). For very thin Al films (5 nm) we simply observe the dependence of the energy gap on the Mo film thickness. Figure 9 shows the density of states through the bilayer. There is absolutely no difference between the density of states in the two materials and it is equal to the density of states in bulk Mo.

## B. Comparison with experiments

In addition to the calculations presented in the previous section a number of calculations were performed which match the specific geometries of STJs that can be measured in the laboratory. The basic parameters used for these calculations are the same as those deduced in section III with exception of the film thicknesses. Figures 10 and 11 show the correspondence between experiment and theory for a series of seven Ta/Al and six Nb/Al STJs with different Al film thicknesses. The Ta and Nb film thicknesses were 100 nm. The correspondence between the model prediction and the experimental derived energy gap is very good, when one considers the large uncertainties in model parameters and the actual junction characteristics, such as film thicknesses and interface characteristics. Note interface parameters were deduced from the experimental data for Ta with 55 nm of Al and Nb with 30 nm of Al. Hence the agreement in the model and measurements in figures 10 and 11 at these values of Al thickness.

For the case of V/Al bilayers only two values are currently available. Besides the value of a bilayer with 100 nm of V and 8nm of Al for which the interface parameters were determined, we also have a value for an STJ with 30 nm of Al. For this STJ the energy gap is in good agreement with the model with a measured gap of 0.53 meV, compared to the predicted value of 0.521 meV.

## V. Conclusions

A model for the energy gap in transition metal/aluminium bilayers has been developed and validated with experimental data from three different lay-ups: Ta/Al, Nb/Al and V/Al. Based on this model it is now straightforward to consider any bilayer geometry of these materials. With certain assumptions the model has been generalised to any transition metal/aluminium bilayer and applied successfully to the Mo/Al system currently under development as a Superconducting Tunnel Junction.

**Tables**

|    | $RRR_b$ | $b_t$ | $T_{c,0}$ | $N\nu$ | $\lambda_{300}$ | bulk $\Delta_g$ | $\xi_0$ |
|----|------|------|-------|-------|--------|--------|------|
|    | /    | nm   | K     | /     | nm     | meV    | nm   |
| Al | 5    | 0.2  | 1.2   | 0.175 | 10.3   | 0.179  | 1600 |
| Ta | 27   | 0.33 | 4.5   | 0.25  | 3.3    | 0.7    | 90   |
| Nb | 90   | 0.25 | 9.3   | 0.35  | 2.5    | 1.39   | 38   |
| V  | 11   | 0.24 | 5.4   | 0.24  | 3.9    | 0.82   | 45   |
| Mo | 247  | 0.315| 0.915 | 0.16  | 16.6   | 0.140  | 708  |

*Table 1: Basic parameter values for all materials*

|                       |     | Ta film thickness (nm) |       |       |       | Nb film thickness (nm) |       |       |       | V film thickness (nm) |       |       |       |
|-----------------------|-----|-------|-------|-------|-------|-------|-------|-------|-------|-------|-------|-------|-------|
|                       |     | 5     | 20    | 50    | 100   | 5     | 20    | 50    | 100   | 5     | 20    | 50    | 100   |
| Al film thickn (nm)   | 5   | 0.305 | 0.561 | 0.651 | 0.659 | 0.755 | 1.098 | 1.23  | 1.276 | 0.534 | 0.682 | 0.724 | 0.739 |
|                       | 20  | 0.291 | 0.489 | 0.559 | 0.557 | 0.491 | 0.78  | 0.923 | 0.983 | 0.414 | 0.536 | 0.576 | 0.587 |
|                       | 50  | 0.269 | 0.408 | 0.454 | 0.45  | 0.339 | 0.545 | 0.644 | 0.686 | 0.319 | 0.401 | 0.428 | 0.435 |
|                       | 100 | 0.241 | 0.328 | 0.355 | 0.361 | 0.267 | 0.407 | 0.47  | 0.498 | 0.263 | 0.317 | 0.334 | 0.340 |

*Table 2: Energy gap (meV) as a function of film thicknesses for Ta/Al, Nb/Al and V/Al bilayers.*

|                        |     | Mo film thickness (nm) |       |       |       |
|------------------------|-----|-------|-------|-------|-------|
|                        |     | 5     | 20    | 50    | 100   |
| Al film thickness (nm) | 5   | 0.1   | 0.127 | 0.135 | 0.136 |
|                        | 20  | 0.117 | 0.130 | 0.136 | 0.137 |
|                        | 50  | 0.134 | 0.134 | 0.138 | 0.138 |
|                        | 100 | 0.148 | 0.139 | 0.140 | 0.139 |

*Table 3: Energy gap (meV) as a function of film thicknesses for Mo/Al bilayers*

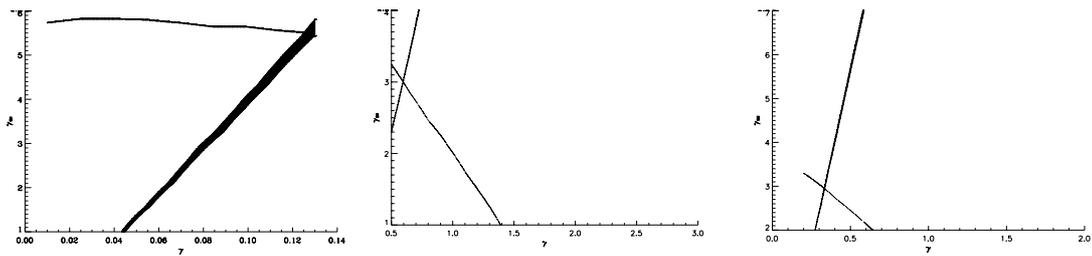

**Figure 1:** The two lines represent the possible combinations of interface parameters for which the model agrees with the experimentally determined critical temperature or energy gap. The intersection of these lines gives the unique pair of interface parameters for this junction. **(a)** Ta/Al STJ with 100 nm of Ta and 55 nm of Al. $\gamma=0.13$ and $\gamma_{BN}=5.5$. **(b)** Nb/Al STJ with 100nm of Nb and 30 nm of Al. $\gamma=0.595$ and $\gamma_{BN}=3$. **(c)** V/Al STJ with 100 nm of V and 8 nm of Al. $\gamma=0.33$ and $\gamma_{BN}=2.95$.

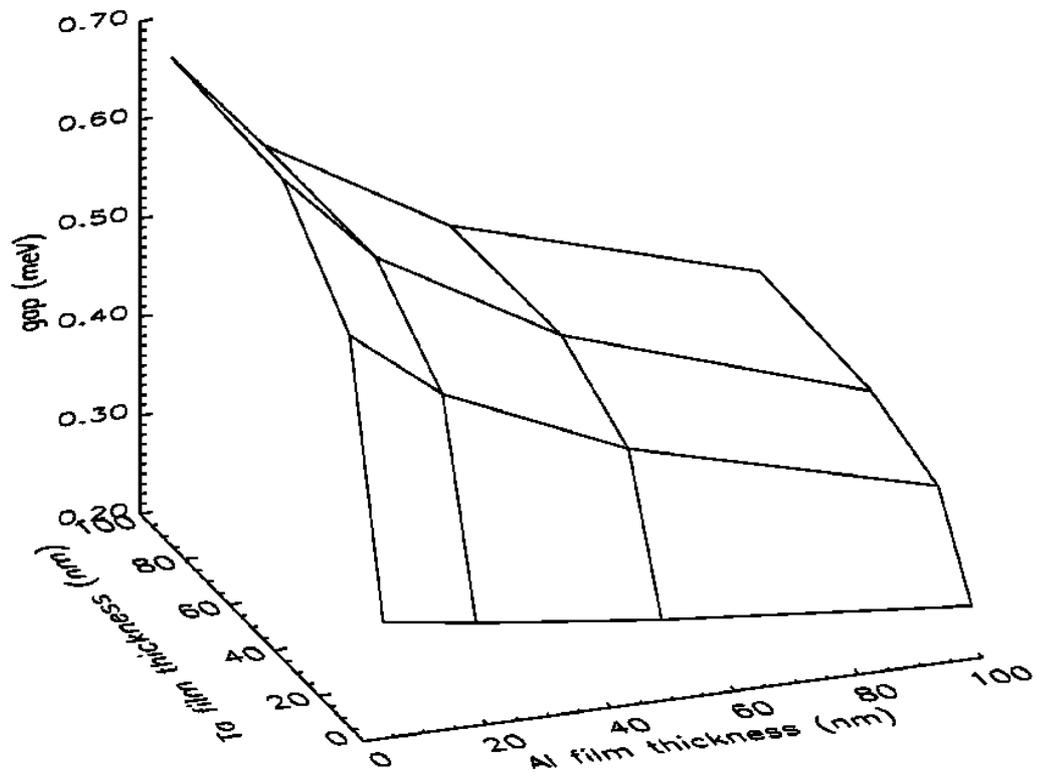

**Figure 2:**
Overall energy gap in a Ta/Al bilayer as a function of Ta and Al film thicknesses.

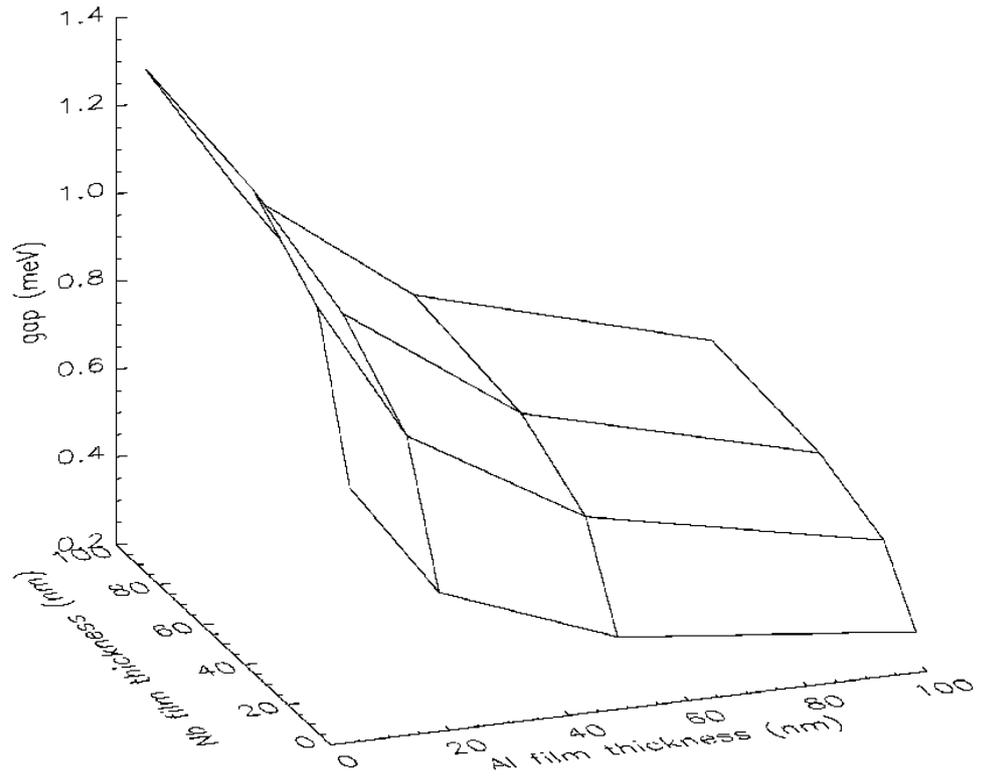

**Figure 3:**
Overall energy gap in a Nb/Al bilayer as a function of Nb and Al film thicknesses.

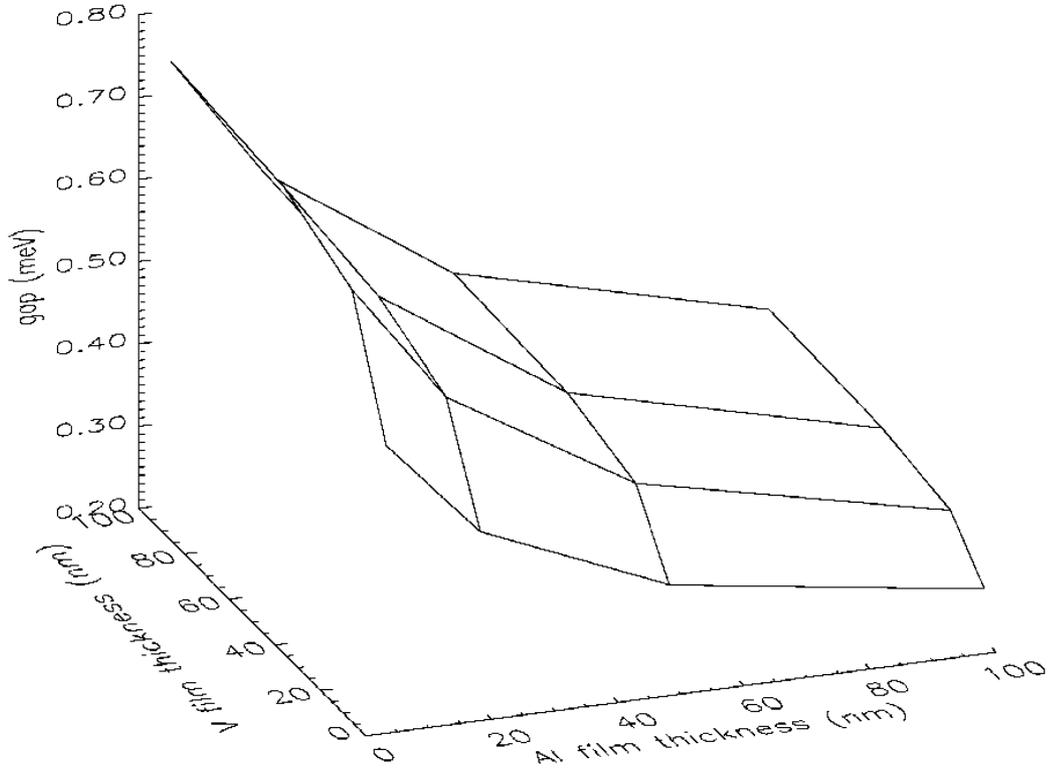

**Figure 4:**
Overall energy gap in a V/Al bilayer as a function of V and Al film thicknesses.

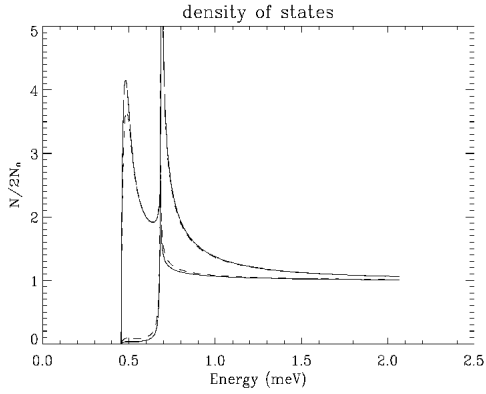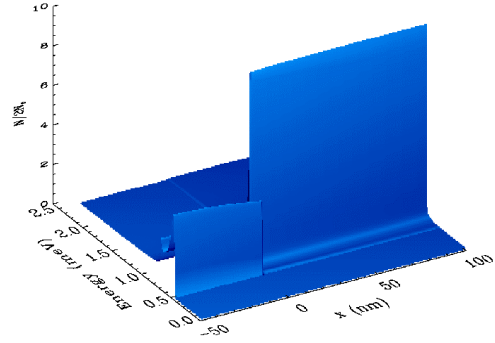

**Figure 5:**

Calculated density of states through a Ta/Al bilayer with 100 nm of Ta and 50 nm of Al. Here **(a)** represents the density of states at four key positions in the bilayer. The dashed lines represent the density of states at the Ta-Al interface, in the Ta and in the Al, while the solid lines provide the density of states at the free interfaces in both films. Figure 5**(b)** illustrates the evolution of the density of states surface as a function of both the position through the whole bilayer and energy.

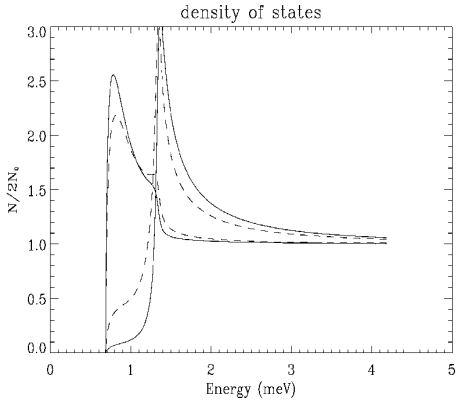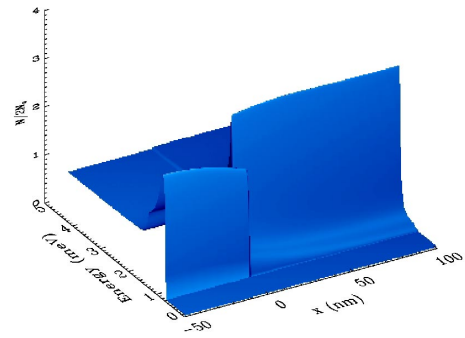

**Figure 6:**

Calculated density of states through a Nb/Al bilayer with 100 nm of Nb and 50 nm of Al. Here **(a)** represents the density of states at four key positions in the bilayer. The dashed lines represent the density of states at the Nb-Al interface, in the Nb and in the Al, while the solid lines provide the density of states at the free interfaces in both films. Figure 6**(b)** illustrates the evolution of the density of states surface as a function of both the position through the whole bilayer and energy.

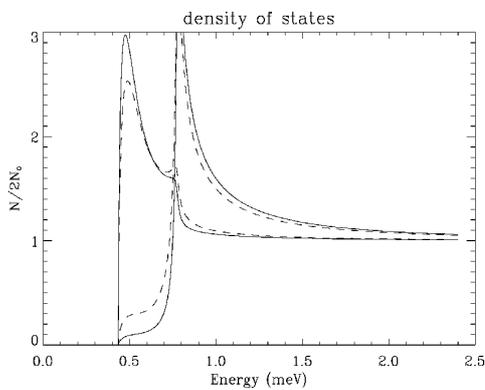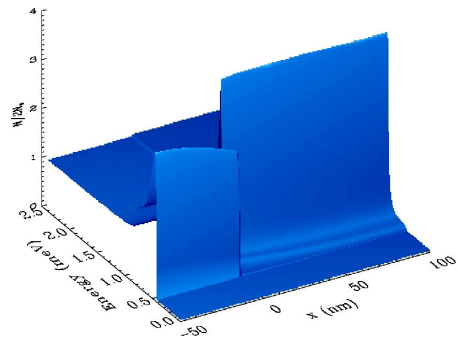

**Figure 7:**

Calculated density of states through a V/Al bilayer with 100 nm of V and 50 nm of Al. Here **(a)** represents the density of states at four key positions in the bilayer. The dashed lines represent the density of states at the V-Al interface, in the V and in the Al, while the solid lines provide the density of states at the free interfaces in both films. Figure 7**(b)** illustrates the evolution of the density of states surface as a function of both the position through the whole bilayer and energy.

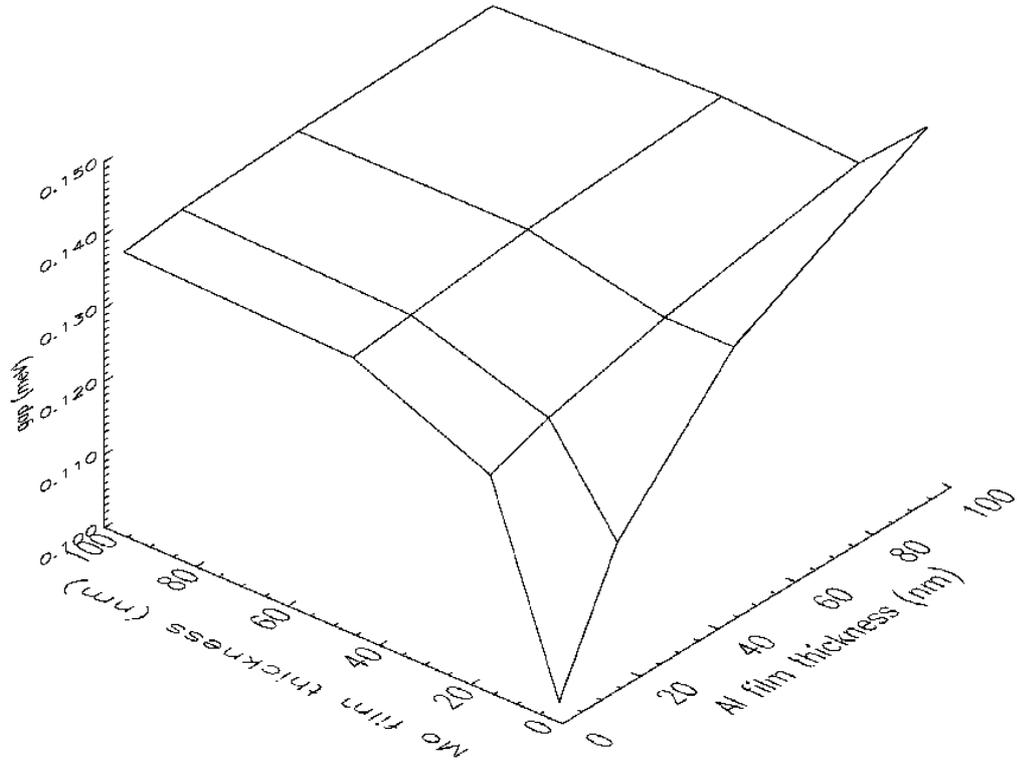

**Figure 8:**
Overall energy gap in a Mo/Al bilayer as a function of Mo and Al film thicknesses.

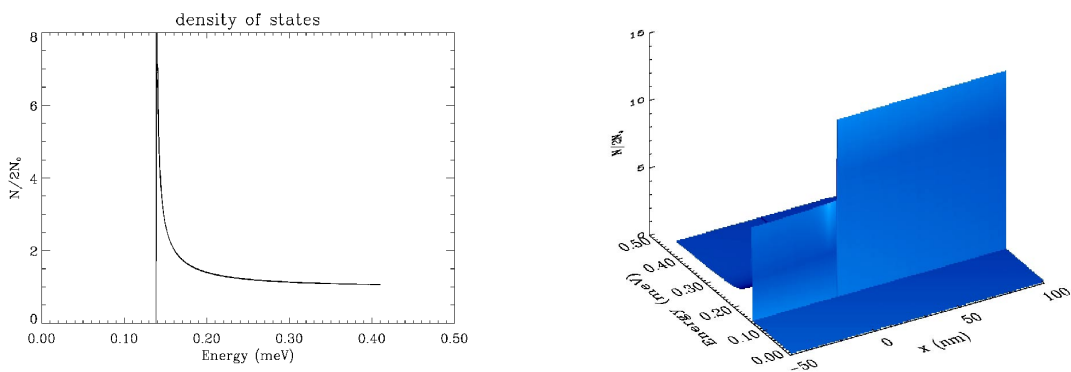

**Figure 9:**
Calculated density of states through a Mo/Al bilayer with 100 nm of Mo and 50 nm of Al. Here **(a)** represents the density of states at four key positions in the bilayer. The dashed lines represent the density of states at the Mo-Al interface, in the Mo and in the Al, while the solid lines provide the density of states at the free interfaces in both films. Figure 7**(b)** illustrates the evolution of the density of states surface as a function of both the position through the whole bilayer and energy.

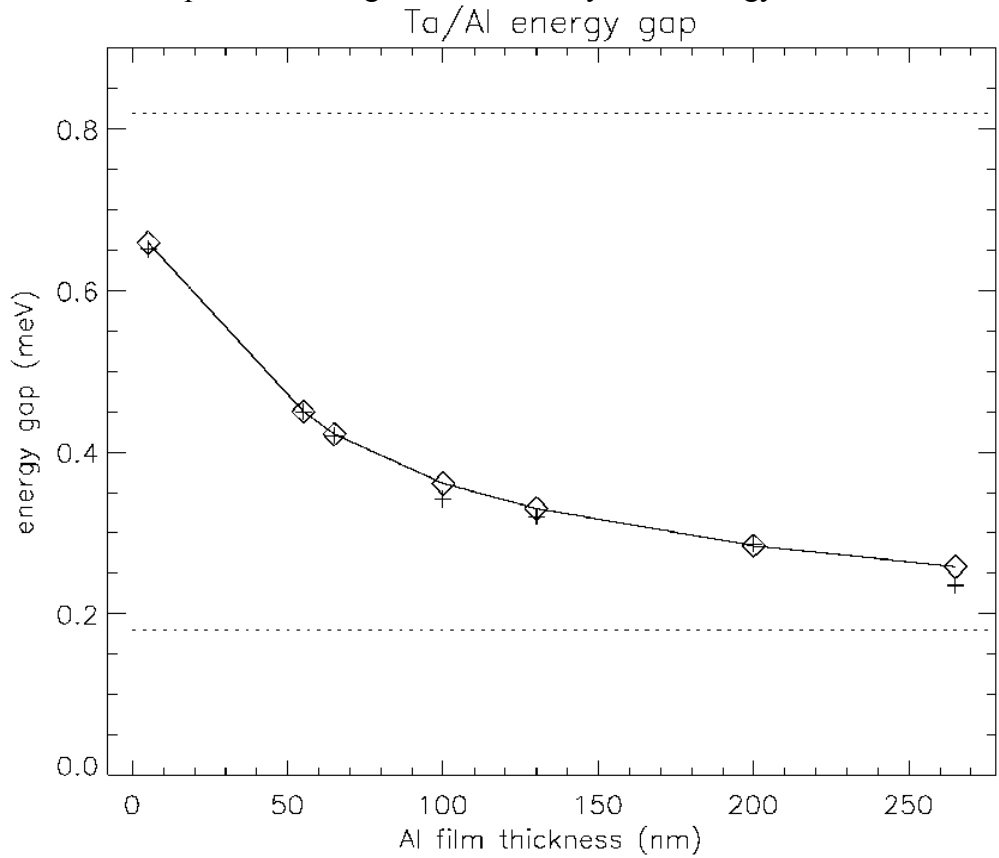

**Figure 10:**
Correspondence between experiment and theory for a series of seven Ta/Al STJs with different Al film thicknesses. The Ta film thickness is 100 nm. The diamonds represent the values calculated with the model, whereas the crosses represent the experimental values. The dashed lines represent the energy gap in bulk Ta and Al.

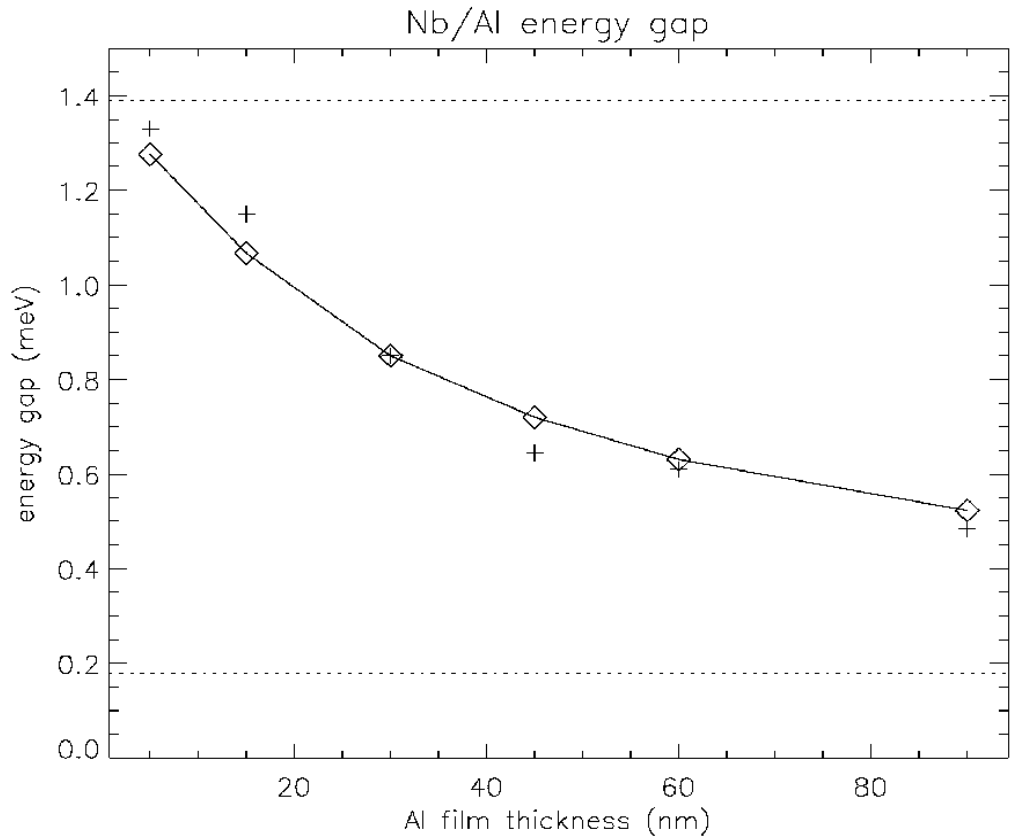

**Figure 11:**
Correspondence between experiment and theory for a series of six Nb/Al STJs with different Al film thicknesses. The Nb film thickness is 100 nm. The diamonds represent the values calculated with the model, whereas the crosses represent the experimental values. The dashed lines represent the energy gap in bulk Nb and Al.

**References**

[1] A. Peacock, Physica B, 263-264 (1999) 595.
[2] K.D. Irwin, G. C. Hilton, D.A. Wollman, J.M. Martinis, J. Appl. Phys., 83 (1998) 3978.
[3] A. Poelaert, Ph.D. thesis, University of Twente, Enschede, The Netherlands, (1999).
[4] K.D. Usadel, Phys. Rev. Lett., 25 (1970) 507.
[5] M. Yu. Kupriyanov and V.F. Lukichev, Zh. Eksp. Teor. Fiz., 73 (1977) 299 [Sov. Phys. JETP, 41 (1977) 960].
[6] L.N. Cooper, Phys. Rev. Lett., 6 (1961) 689.
[7] D. Movshovitz and N. Wiser, Phys. Rev. B, 41 (1990) 10503.
[8] E. H. Sondheimer, Phys. Rev., 80 (1950) 401.
[9] R. Meservey and B. B. Schwartz, Superconductivity Vol. 1 (R.D. Parks, eds.), Marcel Dekker Inc., New York, 1969, p. 126.
[10] C. Reale, Physics Letters, 50A (1974) 53.
[11] M. Gutsche, H. Kraus, J. Jochum, B. Kemmather, G. Gutekunst, Thin Solid Films, 248 (1994) 18
[12] A. Zehnder, Ph. Lerch, S.P. Zhao, Th. Nussbaumer, E.C. Kirk, H.R. Ott, Phys. Rev. B, 59 (1999) 8875.
[13] P.N. Dmitriev, A.B. Ermakov, A.G. Kovalenko, V.P. Koshelets, N.N. Iosad, A.A. Golubov, M. Yu Kupriyanov, IEEE Trans. Appl. Supercond., Vol. 9 No. 2 (1999) 3970.